# Interaction between a Water Molecule and a Graphite Surface

D.J. Wallace Geldart[1,2], I. Wayan Sudiarta[1], Glen Lesins[1], and Petr Chylek[1,3]

[1] Department of Physics and Atmospheric Science, Dalhousie University, Halifax, NS B3H 3J5 Canada
[2] School of Physics, University of New South Wales, Sydney, NSW 2052, Australia
[3] Los Alamos National Laboratory, Los Alamos, NM 87545, USA

Abstract

The interaction energy between a water molecule and graphitic structured clusters terminated by hydrogen atoms is analyzed by ab initio methods and decomposed into electrostatic, induction, Pauli repulsion, and correlation energy contributions. Contributions to the energy which are due solely to the perimeter of the clusters are identified. These can be isolated and discarded which greatly simplifies the problem of extrapolation to the large cluster limit. The remaining terms are intrinsic to the interaction of a water molecule with real graphitic layers and an explicit analytical form is given for the potential energy surface. The minimum energy configuration is found to have both hydrogen atoms of the water molecule pointing symmetrically away from the graphitic plane. The electronic interaction in this mode is -16.8 ±1.7 kJ/mol for water-graphite and the zero point energy is estimated as 1.3 kJ/mol.

## I. Introduction

The interaction of water molecules with a graphitic surface is of fundamental importance in many areas of science and technology. Among the examples of current interest are cloud droplet formation by condensation of water molecules on graphitic structured soot aerosols and applications of carbon nanotubes in aqueous environments or as sensor devices. We will describe briefly the atmospheric science application since it is less well known.[1]

Water molecules in the atmosphere interact with graphitic structured soot aerosols in the process of cloud droplet formation. Freshly produced soot particles from fossil fuels and biomass combustion are generally believed to be hydrophobic.[2,3] With time the soot aerosols can become partially oxidated, becoming hydrophilic with the result that aged soot aerosols can act as condensation nuclei for cloud droplets.[4]

The strength of the interaction between a water molecule and the carbon atoms of a carbonaceous aerosol is one of the most fundamental factors in determining adsorption leading to possible wetting and film formation. A quantitative modeling study of the conversion of soot from a hydrophobic to hydrophilic state requires an accurate determination of this interaction energy. Discussion in this paper is limited to an ideal surface with no defects and which we assume to be comprised of nanoscale covalently bonded graphitic structures. The influence of surface defects and ionization sites will not be considered in this paper. Ultimately all information regarding interaction strengths is to be incorporated into effective potential energy surfaces from which models can then be developed for use in molecular dynamics and studies of kinetics.

Experimental information is sparse but ab initio methods have been applied to the computation of the energy of interaction of a water molecule with planar arrays of carbon atoms. The single graphitic layer was modeled by a sequence of polycyclic aromatic hydrocarbons of increasing cluster size; that is, planar arrays of covalently bonded fused benzene rings terminated by hydrogen atoms at the outer perimeter. The water molecule is located above the center of the cluster and as the cluster size increases the interaction energy of water with the cluster approaches the limiting case of water interacting with an extended graphitic layer.

The most extensive computations are those of Feller and Jordan[5] who calculated electronic interaction energies to second order in Møller-Plesset perturbation theory (MP2) for a water molecule interacting with single layer graphitic structured clusters (up to size $C_{96}H_{24}$). An approximate method was used to correct for basis set superposition errors (BSSE) and to extrapolate to the graphitic limit (graphene). Feller and Jordan[5] used very large basis sets (augmented with diffuse functions) on multiple atomic centers. Nevertheless the BSSE were not small. Their estimate of the electronic interaction energy at the MP2 level of a water molecule interacting with a single extended graphitic layer was -24.3 ± 1.7 kJ/mol. Other ab initio computations will be discussed in Section IV.

Karapetian and Jordan[6] have indicated that uncertainties and error bars are particularly difficult to determine when BSSE corrections are not small and that the correction procedures given in Ref. 5 may overestimate the magnitude of the water-graphite interaction energy. In fact, the estimate of -24.3 ± 1.7 kJ/mol is considerably larger than the interaction energies, -6.7 kJ/mol to -10.0 kJ/mol, found in numerical simulations to reproduce the range of experimental results for the contact angle of water on graphite.[7] From the analysis of Werder et al.,[7] an interaction energy of -24.3 kJ/mol gives a contact angle $\Theta=0°$ for a water droplet on a graphitic surface and the graphitic surface would be strongly hydrophilic in disagreement with experiment.

These ab initio calculations by Møller-Plesset perturbation theory for large cluster systems face two major extrapolations. The first is the extrapolation to the complete basis set limit for each specific cluster of the chosen sequence. Achieving sufficient size of basis set for any given large cluster, while maintaining good quality of the basis functions and including corrections for the BSSE, is an ongoing challenge.[8] The second is the subsequent extrapolation of the sequence of such clusters to the large cluster limit (graphene). The combination of these two extrapolations poses significant computational demands on ab initio studies of these weakly bound complexes.

Plane-wave-based density functional theory (DFT) methods using periodically replicated simulation cells are computationally efficient. Unfortunately the explicit functional representations available at present for the correlation contribution to the energy are not sufficiently accurate to describe the weak van der Waals (dispersion) interactions which are a major source of binding in the water-graphite system. Progress in including the weak van der Waals interactions in a DFT framework, whether empirically[9] or by explicit functionals,[10] is being made. However these formulations are still in development and there are uncertainties for applications to complexes in which dispersion forces are a major source of binding. Consequently the calculations of correlation energies in this paper are made at the MP2 level of ab initio theory.

The purpose of this paper is to develop a new strategy for computing the binding energy of a water molecule to an extended graphitic surface. We begin by giving a physical picture based on pair-wise interactions of the various contributions to the total interaction energy of a water molecule with a cluster of fused benzene rings. With this interpretation, we can identify contributions to the computed interaction energies which are due solely to the charge distribution caused by the saturating hydrogen atoms at the perimeter of the cluster. These perimeter terms can be large for finite clusters but their contribution is exactly zero in the large cluster limit (graphene). Consequently, accuracy of the extrapolation from small cluster data to the large cluster limit is greatly improved by isolating the boundary contributions in finite cluster data and removing them before extrapolation. All remaining terms are intrinsic water-carbon interactions and it is only these terms that contribute in the limit of a real extended graphitic surface. This analytical separation of intrinsic water-carbon interactions from those due to the cluster perimeter effectively removes the problem of numerical extrapolation to the large cluster limit. It is then possible to give a much improved determination by ab initio methods of the structure and binding energy of water on graphene and on graphitic multi-layers.

In Section II the electrostatic, induction, and Pauli (exchange) repulsion contributions are obtained from computations of the Hartree-Fock (HF) electronic energy over a grid of positions and orientations for a water molecule interacting with $C_{24}H_{12}$ and $C_{54}H_{18}$. The MP2 correlation interaction energy of water-benzene is computed using a sequence of large basis sets also over a range of positions and orientations of the water molecule. The distance dependence is found to be well described by dispersion interactions. Parameters for all pair-wise interactions were determined by least squares fitting to this combination of HF and dispersion energies. The extrapolation to the large cluster limit is then accomplished simply by dropping all perimeter interaction energies and retaining only the intrinsic water-carbon interactions. Results for the extrapolated binding energy of water on graphene and on graphitic multi-layers determined in this way are given in Section III. Discussion and comparison with previous work are given in Section IV and Section V is a brief summary and conclusions.

## II. Analysis of Contributions to Interaction Energy

### A. Introduction

The electronic contribution to the ground state energy of a system is the sum of the Hartree-Fock (HF) energy $E_{HF}$ and the correlation energy $E_C$, computed for fixed positions of all nuclei

$$E = E_{HF} + E_C \qquad (1)$$

The total ground state energy is the sum of the electronic energy and the zero point vibrational energy of the nuclei $E_{ZP}$. Unless specified otherwise, energies in this paper refer to the electronic contributions. The interaction energy $\Delta E$ of a water molecule with the cluster is the energy of the total complex minus the energies of the fully separated water molecule and cluster.

The HF interaction energy $\Delta E_{HF}$ is computed by the standard self-consistent average field method. Contributions to $\Delta E_{HF}$ are interpreted as the electrostatic energy (ES) due to fixed charge distributions (permanent multipoles) in the water and cluster, induction energy and the short range Pauli repulsion energy. Charge transfer energy is small and is neglected. The correlation interaction energy $\Delta E_C$ is computed by MP2 perturbation theory and is interpreted as the van der Waals dispersion interaction. All ab initio calculations were done using the Gaussian98[11] and Gaussian03[12] packages and the BSSE is corrected using the counterpoise (CP) method.[13]

It is very important to account for the contributions to $\Delta E$ due to the terminating hydrogen atoms of the finite cluster. The dominant contribution to this perimeter energy comes from the ES part of $\Delta E_{HF}$ and falls off slowly as a power law (not exponentially) as the cluster size increases. This ES perimeter energy is a real and important aspect of $\Delta E$ for any finite cluster of computationally feasible size but it is spurious from the point of view of the large cluster limit where only the intrinsic water-carbon interactions of the graphite sheet are present. Before attempting any extrapolation from small cluster data to the large cluster limit, it is necessary to remove the long range ES perimeter energy. Fortunately the HF energy converges rapidly with respect to increasing size of basis set so quantitatively reliable ab initio HF results are obtainable (see following subsection).

The correlation energy contains the intrinsic dispersion energy between water and carbon atoms and the dispersion energy between the water and the perimeter hydrogens. Both can be extracted from MP2 computations of the correlation energy of the water-benzene complex. However, the MP2 correlation energy converges slowly with respect to increasing size of basis sets, in contrast to the HF energy so we have considered two ways to estimate the dispersion energy parameters. We first compute $\Delta E_C$ using large basis sets to get good initial estimates. We then scale this computed MP2 correlation energy to reproduce the experimental result for the dissociation energy of the water-benzene complex. This requires a computation of the zero point contribution to $\Delta E$. The dispersion energy parameters obtained in these two ways agree to about 10%.

## B. Hartree-Fock energy

The ab initio HF interaction energies of a water molecule and coronene ($C_{24}H_{12}$) and of a water molecule and circumcoronene ($C_{54}H_{18}$) were computed. These two clusters were chosen in order to verify that the parameters describing the intrinsic water-carbon HF interaction energy are independent of cluster size. All calculations used a fixed carbon-carbon (C-C) bond length of 1.421 Å[14,15] and a fixed carbon-hydrogen (C-H) bond length of 1.084 Å[16] are used for coronene and circumcoronene. For the water molecule, the O-H bond length is 0.9572 Å and the $\angle$(H-O-H) bond angle is 104.52°.[17] Other parameters are optimized.

Our main interest is the interaction of water with an infinite graphitic planar surface. Since the van der Waals interaction between a water molecule at a equilibrium separation of about 3 Å from a graphite surface is weak we have used fixed bond lengths of graphite and have neglected any small surface relaxation. Similarly, only one-layer cluster models are used for determining the parameters of the interaction energy. The effect of adding another layer is negligible because of the large interplanar separation in graphite (about 3.4 Å). Although the resulting interaction parameters are obtained for single-layer systems this procedure yields an accurate water-carbon potential energy which can be applied to any number of carbon atoms arranged in one or more layers.

The electrostatic component of the computed HF energy is computed by the interaction of the partial charges of an isolated water molecule with the electrostatic potential due to an isolated graphitic cluster.

$$\Delta E_{ES} = \sum_{i=1}^{3} q_i V_i \qquad (2)$$

$\{q_i\}$ are the partial charges of water using the MP2 computed electron density and Merz-Kollman-Singh (MKS) method.[18,19] $V_i$ is the electric potential at the position of the $i^{th}$ atom of water computed using the MP2 computed electron density of the isolated graphite cluster. Equation (2) contains the interaction between the water dipole moment and the quadrupole moment associated with the carbon atoms as well as the interaction of the water molecule with the perimeter. The interaction between the water dipole moment and the carbon quadrupole moments is a real contribution to the interaction energy. This contribution was independently computed using the partial charges of water and the experimental quadrupole moment of carbon given in Ref. 20. It was found to be less than 0.4 kJ/mol at equilibrium separations and is subsequently neglected.

After subtracting Eq. (2) from the computed HF energy, the intrinsic water-carbon repulsion and induction energy ($\Delta E_{R+Ind}$) remains. It is generally known that the induction energy is much smaller than the repulsion energy for a short separation distance between water and the cluster. We have confirmed that the energy $\Delta E_{R+Ind}$ falls off exponentially as a function of the distance between the water and the cluster plane (not shown), so the functional form for repulsion energy can be represented by

$$\Delta E_R(r) = \sum_{i,j} A_{ij} \exp(-B_{ij} r_{ij}) \qquad (3)$$

The subscript $i$ labels the atoms in the water molecule while $j$ labels the carbons of graphite cluster, and $r_{ij}$ is the distance between atoms $i$ and $j$.

The induction energy has two contributions. The interaction of the induced polarization of the cluster with the static electric field of the water molecule is

$$\Delta E_{ind}(r) = -\sum_{i,j} \frac{C_{ind,ij}}{r_{ij}^6} \qquad (4)$$

The interaction of the induced polarization of the water molecule with the static electric field of the cluster's quadrupole moment has a similar form but can be neglected since the terms vary as $1/r_{ij}^8$ and also because the quadrupole moment of the cluster is small.

Combining Eqs. (3) and (4), we have,

$$\Delta E_{R+Ind}(r) = \sum_{i,j} A_{ij} \exp(-B_{ij} r_{ij}) - \frac{C_{ind,ij}}{r_{ij}^6} \qquad (5)$$

In order to see the convergence of $\Delta E_{R+Ind}$ as a function of cluster size, we have computed HF energies and electrostatic energies for three basis sets and two graphitic clusters ($C_{24}H_{12}$ and $C_{54}H_{18}$). The results are given in Table I and show that the electrostatic energy and consequently the HF energy depend strongly on cluster size. When the electrostatic energy is subtracted from the HF energy, there is good convergence with cluster size. Therefore it can be concluded that using $C_{24}H_{12}$ to compute $\Delta E_{R+Ind}$ is sufficient.

We make a least squares fit to Eq. (5) of the computed $\Delta E_{R+Ind}$ for 16 different distances of water from graphite and two orientations of water (as shown in Fig 1). A sequence of basis sets was used to verify convergence. The parameters for the repulsion energy showed good convergence. However we considered the results for the induction parameters to be less reliable. We attribute this to numerical instability due to interplay between the two induction energy parameters in the fitting procedure. To remove this instability, the number of parameters was reduced by constraining the parameters $C_{ind,ij}$(O-C) and $C_{ind,ij}$(H-C) such that their ratio for all basis sets is the same as the ratio for the cc-pVTZ basis set. The results of this fitting procedure for $C_{24}H_{12}$ are shown in Table II and Figs. 2 and 3. The standard errors in Table II are computed by using fixed values of parameters $B_{ij}$. Note that the $B_{ij}$ coefficient is large enough to justify neglecting the Pauli repulsive contribution from the perimeter hydrogens.

As seen in Table II that the coefficients of repulsion and induction energies show good convergence as the quality of basis sets increases. Therefore, the results for cc-pVTZ basis set will be used for the parameters in Eq. (5).

### C. Correlation energy

It is expected that the correlation interaction energy of the weakly bound water-benzene complex is dominated by the van der Waals dispersion forces. The leading term is due to induced fluctuating dipoles and will be represented as a sum of atomic-like pair interactions

$$\Delta E_{disp} = -\sum_{i,j} \frac{C_{6,ij}}{r_{ij}^6} \quad (7)$$

We will first verify that MP2 correlation energies give the characteristic $1/r^6$ dependence for the interaction between water and benzene and then determine values for $C_6$(O-C), $C_6$(H-C), $C_6$(O-H) and $C_6$(H-H) by least squares fits to the MP2 correlation energy data.

Ab initio MP2/ aug-cc-pVDZ(CP) and MP2/ aug-cc-pVTZ(CP) calculations of the correlation energy for the water-benzene complex were made using the full counterpoise correction. The water molecule was taken to be symmetrically located (mode alpha and beta in Fig. 1) above the center of the carbon ring and the interaction correlation energy was calculated for a sequence of distances between the water molecule and the carbon ring. The interaction correlation energy was found to vary as the inverse sixth power of the O-C distance which confirms the dominance of the dispersion interactions.

The MP2/ aug-cc-pVDZ(CP) and MP2/ aug-cc-pVTZ(CP) correlation energies were then computed for a series of different distances and orientations of the water molecule relative to benzene. The positions and orientations of water were randomly distributed with 50 data points taken over a spatial grid. A least squares fit of this interaction correlation energy data set to Eq. (7) then allows determination of the interatomic $C_6$ dispersion constants together with estimates of their probable error. To avoid instability in the fitting procedure the small parameter $C_6$(H-H) (estimated from Ref. 9 to be 160 kJÅ$^6$/mol), was neglected and the data were fitted using only three parameters. Results of fitting are shown in Fig. 4 and the parameters obtained are given in Table III. Results for the two basis sets were consistent within the estimated error bars.

A variety of procedures were used for the fitting to Eq. (7) to test sensitivity, always with similar results. We verified that the intrinsic water-carbon parameters were not significantly altered by repeating the fits with $C_6$(H-H) fixed at its estimated value.[9] We also examined the effect of including $C_8/r^8$ correction terms in the dispersion energy. The net effect on binding energies was not found to be significant.

Convergence with respect to basis set size is much slower for the correlation energy than for the HF energy. In addition the MP2 level of approximation may overestimate the effect of correlations. In order to obtain another estimate of interaction parameters we adopted an alternative procedure in which the correlation energy is calibrated by experimental results for the dissociation energy of the water-benzene complex. The dissociation energy $D_0$ is defined by

$$\Delta E_{TOTAL} = \Delta E_{HF} + \Delta E_C + \Delta E_{ZP} = -D_0 \quad (6)$$

Experimental measurements by Courty et al. [21] give $D_0 = 10.21 \pm 0.38$ kJ/mol for water on benzene. The various contributions to $\Delta E_{TOTAL}$ on the left hand side of Eq. (6) were obtained as follows.

HF energies have been determined for a sequence of basis sets in order to verify convergence with respect to basis set size. The result for cc-pVDZ, cc-pVTZ and cc-pVQZ basis sets with the full CP correction are -3.43, -3.31 and -3.31 kJ/mol, respectively. This HF value of -3.31 kJ/mol is used to obtain the correlation energy from experimental dissociation energy.

The zero point contribution to the interaction energy has previously estimated by Feller[22] to be 4.18 kJ/mol based on a normal mode analysis with the MP2/aug-cc-pVDZ basis set without CP corrections. We extended this normal mode analysis to CP-corrected energies. The resulting zero point energy for this MP2/aug-cc-pVDZ(CP) calculation is 2.64 kJ/mol. This 40% reduction in zero point energy is due to softer intermolecular interactions and vibrational frequencies with CP corrections.

With this value for the zero point energy and the experimental dissociation energy, the total electronic interaction energy is -12.85 kJ/mol. Subtracting the HF contribution (-3.31 kJ/mol), a semi-empirical value for the correlation interaction energy is obtained as -9.54 kJ/mol. Finally the previously fitted $C_6$s are scaled with a constant multiplicative factor in such a way that the resulting correlation energy matches the correlation energy (-9.54 kJ/mol) obtained from the experimental dissociation energy.

These scaled dispersion parameters give an accurate description of the correlation energy of water interacting with benzene and will also apply to water interacting with fused benzene ring clusters. The same scaled $C_6$(O-C) and $C_6$(H-C) are applicable to the correlation energy of water interacting with graphitic sheets.

### III. Interaction Energy of Water Molecule and Graphitic Layers

The decomposition of the interaction energy in II permits unambiguous identification of contributions due to the perimeter. We emphasize that these terms are very important contributions to the interaction energy of finite clusters but they have exactly zero contribution in the limit of an extended graphitic layer where the water molecule and the perimeter are infinitely separated. Therefore the perimeter energy terms can simply be discarded to describe this limit. The remaining terms are intrinsic to the water-carbon interaction. Combining Eqs. (5) and (7), the final form of the electronic interaction energy between water and graphitic layers is

$$\Delta E = \sum_{i,j} A_{ij} \exp(-B_{ij} r_{ij}) - \frac{(C_{ind,ij} + C_{6,ij})}{r_{ij}^6} \tag{8}$$

The parameters used in this expression are given in Tables II and III.

Using Eq. (8), the interaction energy is easily calculated by direct summation for any configuration of the water molecule and any number of carbon planes. Results for the interaction energy between water and graphitic layers obtained in this way are given in Table IV for three orientations of the water molecule. Mode gamma corresponds to the geometry found by Feller and Jordan[5] where their minimum extrapolated energy has one hydrogen atom of water is pointing towards the graphite surface. Mode alpha is the vibrational average of the mode gamma configuration. The mode beta orientation (Fig. 1) has not been suggested previously as the minimum energy geometry for the water-graphite system. We find that the minimum energy occurs for the mode beta orientation of water (Fig. 5); the two hydrogen atoms point away from the graphitic plane.

The error bars quoted in Table IV are obtained from the distributions computed from Eq. (8) for the electronic interaction energy on the assumption that the probability distributions of the various fitted parameters is Gaussian. The minimum electronic interaction energies of graphite and water in mode alpha and beta configurations are -13.8 ±1.3 and -16.8 ±1.7 kJ/mol, respectively, and for the orientation of water with one hydrogen of the water molecule pointing towards graphite surface (mode gamma) is -12.3 ±0.8 kJ/mol.

Finally we consider the zero point energy due to the intermolecular vibrations of water and the graphitic layer. A normal mode estimate was obtained by fitting a harmonic form around the local minimum in the energy given by Eq. (8) as a function of six intermolecular degrees of freedom. The resulting estimate is 1.3 kJ/mol for the zero point interaction energy of the water-graphene system. This is a smaller correction than for the water-benzene system because the confining effect of the perimeter is absent for water-graphene. The dissociation energy of water from a graphite surface is then 15.5 ±1.7 kJ/mol.

**IV. Discussion**
The minimum energy of a water molecule at a graphite surface occurs when the orientation of water is mode beta (Fig. 5) with the two hydrogen atoms pointing away from the graphitic plane. In contrast, mode alpha or gamma has been found to be stable for water on the strictly finite clusters studied thus far by MP2[5], semiempirical[9], and hybrid[23, 24] methods. This structure is also found in pure HF calculations[25] where the dispersion energy attraction is totally ignored. In this case the attraction responsible for binding of the water molecule to the cluster is the purely electrostatic interaction of the dipole moment of the water molecule with the charge distribution induced in the region of the perimeter by the terminating hydrogen atoms. Of course, this electrostatic component continues to be an important part of the total interaction when the weak (dispersion) correlation energy is added. It is clear from Section II that this electrostatic energy due to the perimeter decreases very slowly (as a power law, not exponentially) as the cluster size increases.

The clusters studied to date are still relatively small and the effect of the perimeter atoms is correspondingly large. The attractive interaction, and hence the stability of the alpha or gamma mode, for these complexes has a large contribution from the perimeter interaction. It is precisely this long range interaction which must be carefully isolated before extrapolation to the large cluster limit. Only then can the stable configuration of water on graphene with intrinsic water-carbon interactions be determined. The issue is whether a hydrogen bond to the graphite surface (modes alpha or gamma) is weaker than bonding of the lone electron pair of the oxygen atom with the $\pi$ electrons of the surface carbon atoms (mode beta). There is a delicate balance of attractive and repulsive energies. We find that the mode beta geometry takes maximum advantage of electron correlation while reducing the Pauli repulsion.

There is other direct evidence of the important role of the perimeter energy in determining the equilibrium structure of water-cluster complexes. Raimondi et al.[26] have calculated the electronic energy and geometry of the water-hexafluorobenzene complex ($H_2O$-$C_6F_6$) using the medium size diffuse 6-31G(d=0.25) basis set and ab initio methods at the MP2 level. Mode beta was found to be the stable structure. Sudiarta and Geldart[27] have compared the energies and structures of sequences of fluorine-terminated versus hydrogen-terminated complexes at the same 6-31G(d=0.25) MP2 level. All fluorine-terminated complexes had mode beta structure and all hydrogen-terminated complexes had mode alpha structure due to the electrostatic boundary energy. The large electronegativity of fluorine caused the partial charges in the vicinity of the perimeter to change sign relative to the hydrogen-terminated complexes. As a result the dipole moment of the water molecule also changes its direction in order to minimize the energy. This reinforces the need to account properly for the perimeter energy prior to attempting an extrapolation to the large cluster limit.

A second very instructive example is given by the interaction between a water molecule and a $C_{60}$ molecule.[28] This spherically symmetrical molecule has no boundary so the boundary energy is automatically zero in this case. The interaction energy has been calculated by Fomina et al.[28] using the ONIOM(MP2:PM3) method. For sufficiently accurate models and basis function sets the stable configuration of the $H_2O$-$C_{60}$ complex is found to have the two hydrogen atoms pointing away from the surface. This is the precise analogue of mode beta in the case of a planar graphitic sheet. The reason for the stability of this configuration for the $H_2O$-$C_{60}$ complex is the same as for water at a planar graphitic layer. The distance between the oxygen atom of the water molecule and the plane of the nearest carbon hexagon (3.09 Å to 3.19 Å) of $C_{60}$ is comparable to the water oxygen-graphite plane distance (about 2.9 Å from Fig. 5) in mode beta.

**V. Summary and Conclusions**

The interaction energy between water molecules and graphitic surfaces is required for many applications of fundamental interest. Determination of an effective potential energy surface is difficult using conventional ab initio methods for several reasons. The weak long-range dispersion forces responsible for binding water on graphite must be treated accurately, corrections for basis state superposition errors are significant for large clusters, and slowly varying contributions to the energy from the perimeter of the clusters must be accounted for when extrapolating to the large cluster limit.

In this paper we developed a procedure for accurate extrapolation to the large cluster limit. The different contributions to the total interaction energy are separated and analyzed. The HF interaction energy of water with fused benzene ring clusters is computed with 6-31 G(d), 6-311 G(d,p), cc-pVDZ and cc-pVTZ basis sets. The correlation interaction energy computed at MP2/ aug-cc-pVTZ(CP) level is found to be well described by dispersion forces. The zero point interaction energy is also computed. This permits a second estimate of the correlation energy by scaling the magnitudes of the dispersion energy $C_6$ parameters so as to reproduce the experimentally observed dissociation energy of water and benzene. This is a modest correction and does not influence the relative stability of modes. Analysis of the decomposition of the interaction energy into electrostatic, induction, Pauli repulsion, and correlation energies allows the relevant parameters to be obtained by fitting to computed data and an explicit analytical form is obtained for the interaction energy. Contributions to the energy which are due solely to the terminating hydrogen atoms at the perimeter of a finite cluster are identified and discarded. The problem of extrapolation to the large cluster limit is removed by this procedure. Interaction energies can then be obtained by focusing computational resources on the best accurate treatment of relatively small clusters. Note that this procedure is not restricted to Møller-Plesset perturbation theory but applies to other ab initio methods.

After removing the boundary energies, the remaining terms are intrinsic to the interaction of a water molecule with a real graphitic layer and the full potential energy surface can be explored. The minimum energy configuration is found to have both hydrogen atoms of the water molecule pointing symmetrically away from the graphitic plane (mode beta). The electronic interaction in this mode is -16.8 ±1.7 kJ/mol and the dissociation energy for water-graphite is -15.5 ±1.7 kJ/mol.

**Acknowledgements**

This work was supported by the Canadian Foundation for Atmospheric Science and Climate (CFCAS) and the Natural Sciences and Engineering Research Council (NSERC) of Canada.

**References**

[1] P. Chylek, S. G. Jennings, and R. Pinnick, in Encyclopedia of Atmospheric Sciences, edited by J. R. Holton (Academic Press, New York, 2003), p. 2093.

[2] E. A. Müller, L. F. Rull, L. F. Vega, and K. E. Gubbins, J. Phys. Chem. **100**, 1189 (1996).

[3] B. Gorbunov, A. Baklanov, N. Kakutkina, H. L. Windsor, and R. Toumi, J. Aerosol Sci., **32**, 199 (2001).

[4] B. Zuberi, K. S. Johnson, G. K. Aleks, L. T. Molina, M. J. Molina, and A. Laskin, Geophys. Res. Lett., **32**, L01807 (2005).

[5] D. Feller and K. D. Jordan, J. Phys. Chem. A **104**, 9971 (2000).

[6] K. Karapetian, J. D. Jordan, in Water in confining geometries edited by V. Buch and J. P. Devlin (Springer-Verlag, Berlin Heidelberg, 2003), p. 139.

[7] T. Werder, J. H. Walther, R. L. Jaffe, T. Halicioglu, and P. Koumoutsakos, J. Phys. Chem. B **107**, 1345 (2003).

[8] F. B. van Duijneveldt, in Molecular Interactions: from van der Waals to strongly complexes edited by S. Scheiner (Wiley, Chichester, 1997), p. 81.

[9] S. Grimme, J. Comput. Chem. **25**, 1463 (2004).

[10] Axel D. Becke and Erin J. Johnson, 2005, J. Chem. Phys. **123**, 154101(2005).

[11] M. J. Frisch et al. Gaussian 98, Revision A.7 (Gaussian, Inc., Pittsburgh PA, 1998).

[12] M. J. Frisch et al. Gaussian 03, Revision B.05 (Gaussian, Inc., Pittsburgh PA, 2003).

[13] S. F. Boys and F. Bernardi, Mol. Phys. **19**, 553 (1970).

[14] 17. Nixon, D. E. and G. S. Parry, J. Phys. C (Solid St. Phys.) **2**, 1732 (1969).

[15] 18. King, H. W. in CRC Handbook of Chemistry and Physics 78$^{th}$ Ed. edited by D. R. Lide (CRC Press, New York, 1997).

[16] 19. Stoicheff, B. P., Cad. J. Phys. **32**, 339 (1954).

[17] 20. W. S. Benedict, N. Gailar and E. K. Plyler, J. Chem. Phys. **24**, 1139 (1956).

[18] B. H. Besler, K. M. Merz, Jr. and P. A. Kollman, J. Comput. Chem. **11**, 431 (1990).

[19] U. C. Singh and P. A. Kollman, J. Comput. Chem. **5**, 129 (1984).

[20] D.B. Whitehouse and A.D. Buckingham, J. Chem. Soc. -Faraday Transactions **89,** 1909 (1993).

[21] A. Courty, M. Mons, I. Dimicoli, F. Piuzzi, M.-P. Gaigeot, V. Brenner, P. De Pujo, and P. Millié. J. Phys. Chem. A **102**, 6590 (1998).

[22] D. Feller, J. Phys. Chem. A **103**, 7558 (1999).

[23] C. S. Lin, R. Q. Zhang, S. T. Lee, M. Elstner, Th. Frauenheim, and L. J. Wan, J. Phys. Chem. B **109**, 14183 (2005).

[24] S. Xu, S. Irle, D. G. Musaev, and M. C. Lin, *J. Phys. Chem. A*, *109,* 9563 (2005).

[25] H. Ruuska, and T. A. Pakkanen. Carbon **41**, 699 (2003).

[26] M. Raimondi, G. Calderoni, A. Famulari, L. Raimondi, and F. Cozzi, *J. Phys. Chem. A*, 107, 772 (2003).

[27] I.W. Sudiarta and D.J.W. Geldart, *J. Phys. Chem. A,* **110** (35), 10501 (2006).

[28] L. Fomina, A. Reyes, P. Guadarrama, and S. Fomine, Int. J. Quantum Chem. **97**, 679 (2004).

[29] G.Schaftenaar and J.H. Noordik, J. Comput.-Aided Mol. Design **14**, 123 (2000).

List of Tables:

Table I. The Hartree-Fock (HF) and electrostatic (ES) interaction energies for water-coronene ($C_{24}H_{12}$), and water-circumcoronene ($C_{54}H_{18}$) for water at z(O-C) = 3.0 Å above the clusters for mode alpha. All energies are in kJ/mol.

| Cluster/Basis | HF | ES | HF-ES |
|---|---|---|---|
| C24H12/3-21G | 8.602 | -7.748 | 16.350 |
| C54H18/3-21G | 12.014 | -4.990 | 17.004 |
| C24H12/6-31G(d) | 11.402 | -5.355 | 16.757 |
| C54H18/6-31G(d) | 13.541 | -3.172 | 16.713 |
| C24H12/6-311G(d,p) | 12.799 | -4.781 | 17.580 |
| C54H18/6-311G(d,p) | 14.999 | -2.714 | 17.713 |

Table II. The parameters of repulsion and induction energies between water and $C_{24}H_{12}$. The results for $C_{ind,ij}$ for 6-31G(d), cc-pVDZ, and 6-311G(d,p) are constrained such that ratio of $C_{ind,ij}$(O-C) to $C_{ind,ij}$(H-C) is the same as the ratio for cc-pVTZ basis set.

| Basis Set | A(O-C) (kJ/mol) | B(O-C) (Å$^{-1}$) | $C_{Ind}$(O-C) (kJ mol$^{-1}$ Å$^6$) | A(H-C) (kJ/mol) | B(H-C) (Å$^{-1}$) | $C_{Ind}$(H-C) (kJ mol$^{-1}$ Å$^6$) |
|---|---|---|---|---|---|---|
| 6-31G(d) | 88267 | 3.344 | 468 | 9010 | 3.140 | 220 |
| cc-pVDZ | 111337 | 3.442 | 344 | 7133 | 3.072 | 162 |
| 6-311G(d,p) | 130904 | 3.513 | 325 | 6716 | 3.048 | 153 |
| cc-pVTZ | 132857 ± 399 | 3.503 | 317 ± 15 | 7482 ± 72 | 3.135 | 149 ± 8 |

Table III. Fit of correlation energy data with three $C_6$ dispersion parameters for aug-cc-pVTZ(CP) and aug-cc-pVTZ(CP) basis sets. $C_6$ is in kJ/mol (Å)$^6$

| Method | $C_6$(O-C) | $C_6$(H-C) | $C_6$(O-H) |
|---|---|---|---|
| MP2/aug-cc-pVDZ | 2833 ± 205 | 75 ± 29 | 900 ± 205 |
| MP2/aug-cc-pVTZ | 2787 ± 213 | 259 ± 42 | 636 ± 259 |
| Scaling of MP2/aug-cc-pVTZ | 2477 ± 146 | 230 ± 29 | 561 ± 201 |

Table IV. The net electronic interaction energies of water and graphitic layers.

| | Interaction Energies (kJ/mol) | | | |
|---|---|---|---|---|
| Number of layers | 1 layer | 2 layers | 3 layers | 4 layers |
| Mode alpha | -12.4 ±1.3 | -13.6 ±1.3 | -13.8 ±1.3 | -13.8 ±1.3 |
| Mode beta | -15.3 ±1.3 | -16.6 ±1.7 | -16.8 ±1.7 | -16.8 ±1.7 |
| Mode gamma | -11.0 ±0.8 | -12.0 ±0.8 | -12.2 ±0.8 | -12.3 ±0.8 |

List of Figures:

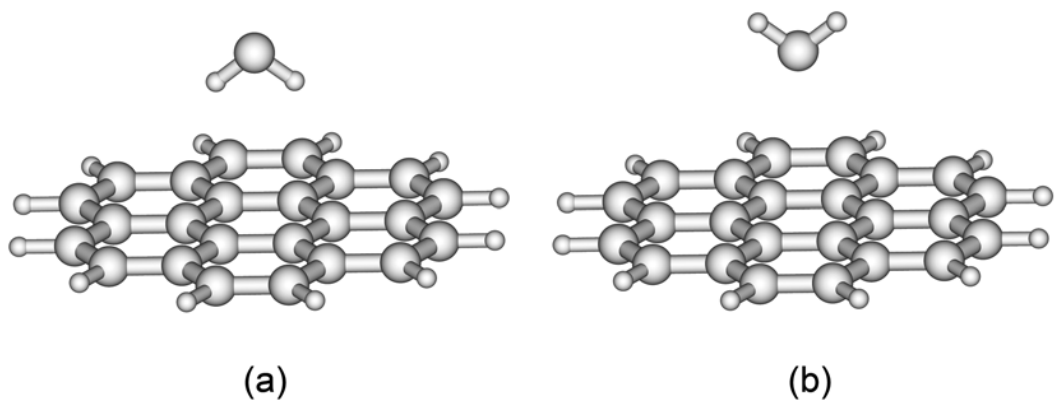

Figure 1. Orientations of water above a $C_{24}H_{12}$ surface used to determine the short range repulsion and the induction energies: (a) mode alpha, and (b) mode beta. This illustration was created using Molden.[29]

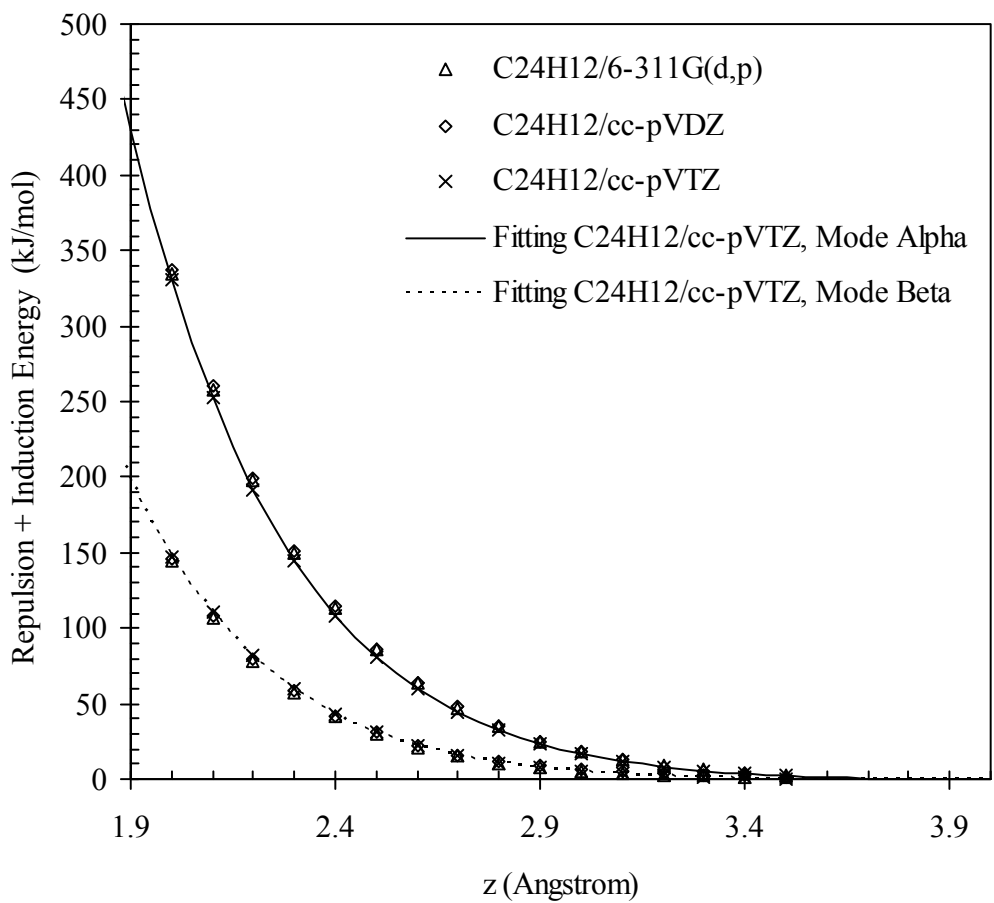

Figure 2. Repulsive + induction interaction energy of water centered over a graphite cluster as a function of distance between the oxygen of water and the cluster calculated by subtracting the electrostatic energy from the HF energy. This energy is computed with 6-311G(d,p), cc-pVDZ and cc-pVTZ basis sets. Note that the upper curve is for water in mode alpha and the lower curve is for water in mode beta.

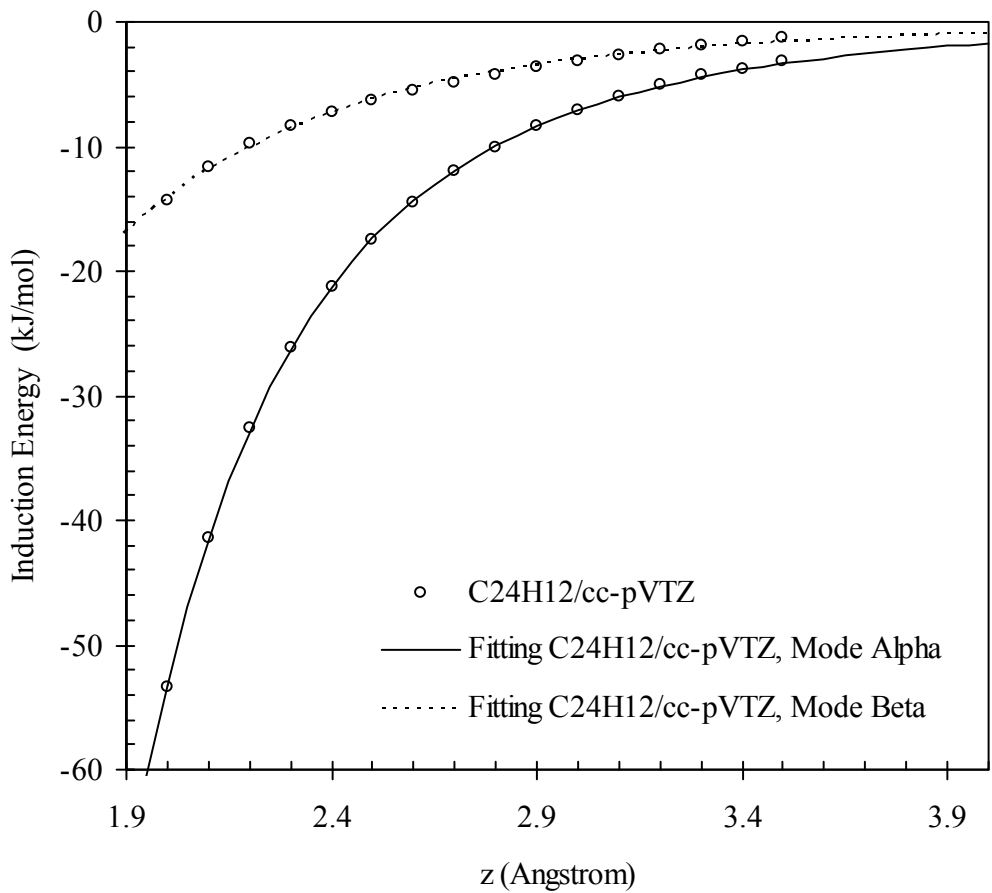

Figure 3. Induction interaction energy of water centered over a graphite cluster as a function of distance between the oxygen of water and the cluster calculated by subtracting the repulsion energy from the HF-ES energy. The repulsion energy is computed using Eq. (3) with fitting parameters in Table II. Note that the upper curve is for water in mode beta and the lower curve is for water in mode alpha.

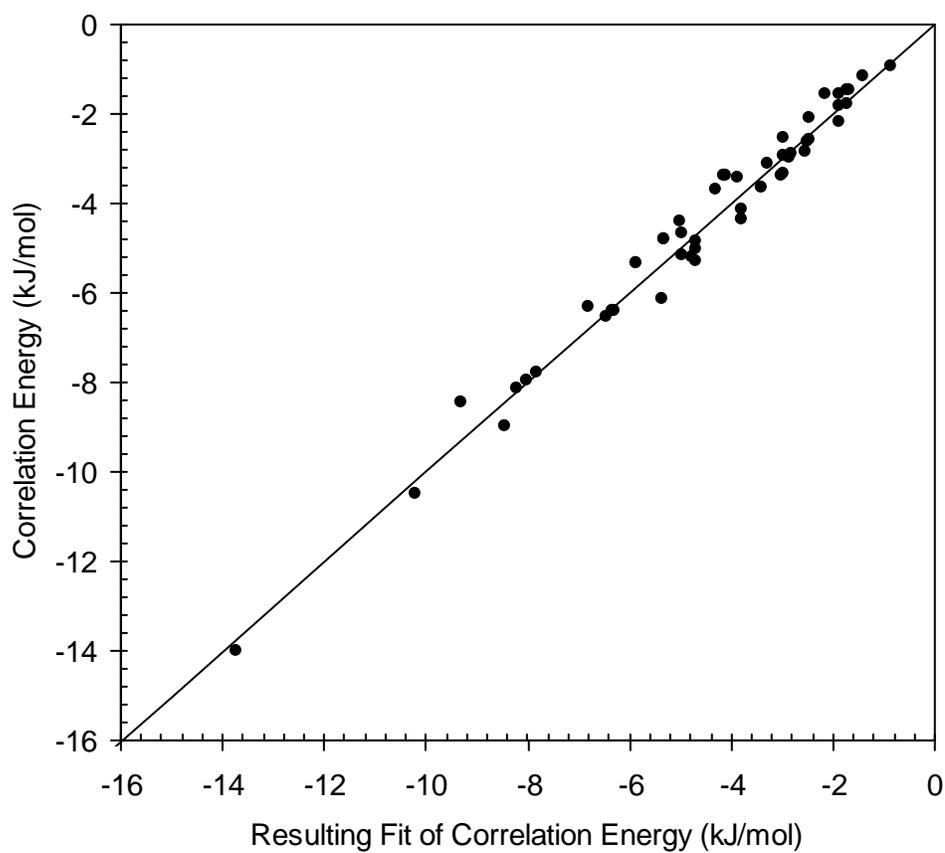

Figure 4. Resulting fit of correlation energies computed using MP2 with aug-cc-pVTZ basis set for 50 different configurations. The resulting parameters are given in Table III.

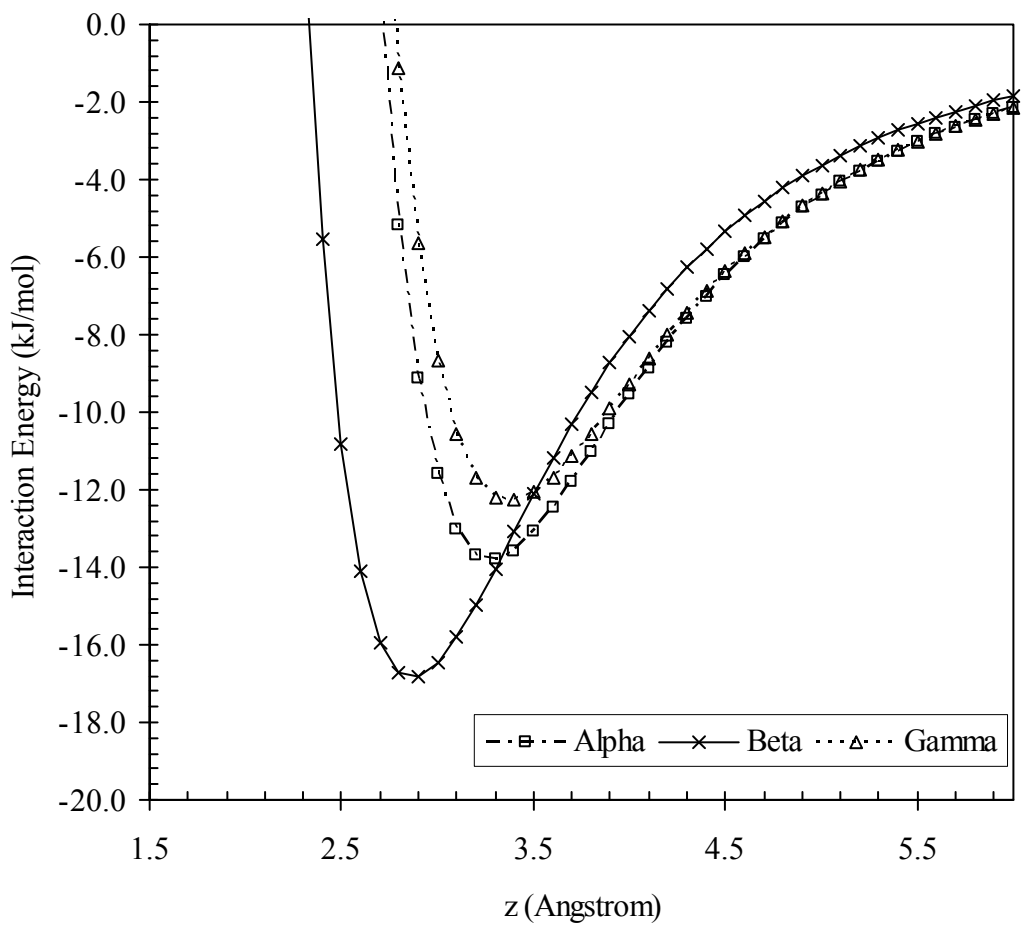

Figure 5. Interaction energy of water molecule with graphite surface for 3 water orientations as a function of the distance between the oxygen of water and graphite surface above the center of a benzene ring. Note that the curve labeled gamma (triangle symbol) is for a water with one hydrogen pointing towards graphite, the curve with square symbol is for a water in mode alpha and the solid curve with the cross symbol is for a water in mode beta.